\documentclass[twocolumn,apjfonts,tighten]{aastex631}
\usepackage{apjfonts}
\usepackage{graphicx}
\usepackage{wrapfig}
\usepackage{amsmath}
\usepackage{multirow}
\usepackage{booktabs}
\usepackage{float}
\usepackage{multirow}

\begin{document} 

\title{Catalog of Integrated-Light Star Cluster Light Curves in TESS} 
\shorttitle{Star Cluster Light Curve Catalog}

\author[0000-0001-6320-2230]{Tobin M. Wainer}
\affiliation{Department of Astronomy, University of Washington, Box 351580, Seattle, WA 98195, USA}
\affiliation{Department of Physics and Astronomy, University of Utah, Salt Lake City, UT 84112, USA}

\author[0000-0001-6761-9359]{Gail Zasowski}
\affiliation{Department of Physics and Astronomy, University of Utah, Salt Lake City, UT 84112, USA}

\author[0000-0002-3827-8417]{Joshua Pepper}
\affiliation{Lehigh University, Department of Physics, 16 Memorial Drive East, Bethlehem, PA, 18015, USA}

\author[0000-0001-6147-5761]{Tom Wagg}
\affiliation{Department of Astronomy, University of Washington, Box 351580, Seattle, WA 98195, USA}

\author[0000-0002-3385-8391]{Christina L. Hedges}
\affiliation{NASA Goddard Space Flight Center, Greenbelt, Maryland, United States}
\affiliation{University of Maryland, Baltimore County, 1000 Hilltop Circle, Baltimore, Maryland, United States}

\author[0000-0002-9831-3501]{Vijith Jacob Poovelil}
\affiliation{Department of Physics and Astronomy, University of Utah, Salt Lake City, UT 84112, USA}

\author[0000-0002-3551-279X]{Tara Fetherolf}
\altaffiliation{UC Chancellor's Fellow}
\affiliation{Department of Earth and Planetary Sciences, University of California Riverside, 900 University Avenue, Riverside, CA 92521, USA}

\author[0000-0002-0637-835X]{James R. A. Davenport}
\affiliation{Department of Astronomy, University of Washington, Box 351580, Seattle, WA 98195, USA}

\author{P. Marios Christodoulou}
\affiliation{Lehigh University, Department of Physics, 16 Memorial Drive East, Bethlehem, PA, 18015, USA}

\author[0000-0002-6401-778X]{Jack T. Dinsmore}
\affiliation{Department of Physics, Massachusetts Institute of Technology, 77 Massachusetts Ave, Cambridge, MA 02139, USA}

\author[0009-0000-5120-1193]{Avi Patel}
\affiliation{Departments of Physics and Astronomy, Haverford College, 370 Lancaster Avenue, Haverford, PA 19041, USA}
\affiliation{Division of Engineering and Applied Science, California Institute of Technology, Pasadena, CA 91125, USA}

\author[0000-0002-7743-9906]{Kameron Goold}
\affiliation{Department of Physics and Astronomy, University of Utah, Salt Lake City, UT 84112, USA}

\author[0000-0001-8203-6004]{Benjamin J. Gibson}
\affiliation{Department of Physics and Astronomy, University of Utah, Salt Lake City, UT 84112, USA}


\correspondingauthor{Tobin M. Wainer}
\email{tobinw@uw.edu}

\begin{abstract}
We present the first integrated light, TESS-based light curves for star clusters in the Milky Way, Small Magellanic Cloud, and Large Magellanic Cloud. We explore the information encoded in these light curves, with particular emphasis on variability. We describe our publicly available package \texttt{elk}, which is designed to extract the light curves by applying principal component analysis to perform background light correction, and incorporating corrections for TESS systematics, allowing us to detect variability on time scales shorter than $\sim$10 days. We perform a series of checks to ensure the quality of our light curves, removing observations where systematics are identified as dominant features, and deliver light curves for 348 previously-cataloged open and globular clusters. Where TESS has observed a cluster in more than one observing sectors, we provide separate light curves for each sector (for a total of 2204 light curves). We explore in detail the light curves of star clusters known to contain high-amplitude Cepheid and RR Lyrae variable stars, and confirm that the variability of these known variables is still detectable when summed together with the light from thousands of other stars. We also demonstrate that even some low-amplitude stellar variability is preserved when integrating over a stellar population. 

\end{abstract}


\section{Introduction}
\label{sec:intro}

Stellar cluster populations serve as fundamental probes of star formation activity and galaxy evolution. Star clusters are the long-lasting remnants of the density peaks in past hierarchically structured star-forming regions \citep[e.g][]{grudic_model_2021}, and thus preserve the characteristics of their local star formation environment well after the event occurred \citep[e.g.,][]{adamo_probing_2015, adamo_star_2020, johnson_panchromatic_2016}. Globular clusters have been used to trace the early formation and assembly history of galaxies \citep[e.g.,][]{kruijssen_fraction_2012}, while younger open clusters are used to infer recent star formation histories \citep[e.g][]{cantat-gaudin_painting_2020}. Because star clusters of different ages represent different epochs in a galaxy’s history, their properties yield valuable information about the extended star formation process throughout galaxies' evolution \citep{johnson_panchromatic_2017}.

One important property of stars, and thus also of stellar clusters, is their intrinsic photometric variability, including both high- and low-amplitude variations driven by coherent pulsations and other physical mechanisms like rotation. A great deal of the foundational work on stellar variability, especially of the low-amplitude variety, has been done in the Milky Way, including the recent revolution in asteroseismic measurements of stars across the Galaxy \citep[e.g.,][]{chaplin_asteroseismic_2013,pinsonneault_second_2018}. High-amplitude pulsational variables such as Cepheids and RR~Lyrae associated with clusters have long been used to measure distances and provide constraints on foreground extinction \citep[e.g.,][]{alonso-garcia_variable_2021}. Famously, it was Cepheid variables that were used to measure the distance to M31 and to argue that the then-called ``island universes'' and ``spiral nebulae'' were in fact other galaxies \citep{hubble_spiral_1929}. Because of the high impact uses of stellar variability, a number of globular clusters in the Milky Way have detailed variable star membership catalogs \citep[e.g.,][]{clement_variable_2001}. 

As in globular clusters, for the vast majority of younger open clusters, the study of variability for member stars has often been focused on high amplitude variables \citep[e.g.,][]{medina_revisited_2021}. However, there has also been a recent revolution in characterizing the low-amplitude stellar variability, yielding results about the underlying physics of these processes. For example, stellar rotation periods have been shown to lengthen with stellar age \citep[e.g.,][]{barnes_ages_2007}, enabling numerous gyrochronology studies of Galactic clusters \citep[e.g.,][]{healy_stellar_2020, gillen_ngts_2020, somers_m_2017, godoy-rivera_stellar_2021}. Also within clusters, TESS data has been used to study the light curves of member stars in order to study the physics of stars with known ages \citep[e.g.,][]{bouma_cluster_2019}.

Another recent revolution in low amplitude stellar variability has been the study of asteroseismology. These studies have shown the ability to measure masses and therefore ages of red giant branch stars \citep[e.g.,][]{pinsonneault_second_2018}, opening avenues for asteroseismological studies in galactic clusters where the age of the cluster is well known. One example being the use of TESS-based asteroseismology to constrain the age of the very young open cluster Melotte~20 \citep{pamos_ortega_determining_2022} using candidate $\delta$~Sct stars. 

Extragalactic time-domain studies have been pivotal to understanding the low-amplitude variability of individual massive stars. Long-term variability studies of M31 and M51 using HST \citep{conroy_complete_2018, soraisam_variability_2020} focused on longer-term variability of bright massive stars, finding ubiquitous variability in the most luminous regions of the color magnitude diagram (CMD), while on the main sequence there is increasing prevalence of variability in later spectral types \citep{mcquillan_rotation_2014}. Studies of short-term variability of massive evolved stars in the Small Magellanic Cloud (SMC) using TESS led to candidates for new classes of pulsating supergiants \citep{dorn-wallenstein_short-term_2019, dorn-wallenstein_short-term_2020}. However, this level of detailed extragalactic analysis has been performed on only the brightest, most massive stars in an extremely small number of galaxies. Building a comprehensive statistical sample of the variability characteristics of stellar populations in different host environments, covering a large range of metallicities, mass and age, will help us better understand the physics of stellar variability, the physical drivers of this variability, and --- crucially --- if all stars of a given age, abundance, temperature and luminosity vary consistently. 

Despite the majority of the universe being unresolved to us, studies of the variability of unresolved stellar populations, i.e., of their integrated light, remain very limited. One analysis of unresolved field populations in M87 \citep{conroy_ubiquitous_2015} characterized the density of long-period variable stars, as a prediction for the age of the population. The main issue with using integrated light for variability encountered in this study was the blending of sources. There have been some efforts to address this issue (in TESS, relevant for this current work) for the light curves of individual resolved stars \citep[e.g.,][]{oelkers_precision_2018,nardiello_psf-based_2020, higgins_localizing_2022}. 

For galaxies outside of the Local Group ($d \gtrsim 1$~Mpc), it is not (yet) possible to resolve individual stars in clusters. Even in large Local Group galaxies ($d \lesssim 1$~Mpc) the main sequence turnoff based ages are only measurable for clusters younger than $\sim$300~Myr \citep{johnson_panchromatic_2016, wainer_panchromatic_2022}. For environments where individual stars can not be resolved, the current standard approach to population analysis is through integrated light methods. While integrated light spectroscopy is a bona fide approach to determining reliable abundances \citep[e.g.,][]{zin_globular_1980, armandroff_spectroscopy_1988}, these methods are highly susceptible to degeneracies in age, stellar mass, and extinction, yielding large systematic uncertainties in deriving cluster properties \citep{krumholz_slug_2019}. Some very interesting efforts have been made to bridge this gap in the semi-resolved regime, such as the use of pixel CMDs \citep{conroy_pixel_2016}. However, these approaches have not been applied to stellar clusters or rigorously tested in multiple galaxies. 

Despite these challenges, determining star cluster properties using their integrated light will remain a vital tool in the future. Even in the JWST era, we will not be able to resolve individual stars for clusters outside the local Universe. In these regimes, integrated light methods serve as the only tool at our disposal to study these systems. The development of new methods of integrated light measurements that return reliable property estimates with realistic error bars was identified by the review article of \citep{krumholz_star_2019} as one of the most pressing needs in stellar cluster research. 

This present work uses common integrated light methodology in a relatively new manner, studying the time series of aperture integrated photometry. Specifically, we extract integrated light curves of star clusters, which remain unexplored. The term `integrated' refers to the method in which we are extracting blended photometry from an aperture, a practice common in the literature. Our approach does this extraction at many time steps to generate light curves, and we evaluate how the integrated light changes through time. Throughout the rest of this paper, we will refer to these light curves as `integrated light curves'. 

Just as a small number of (unresolved, bright) stars can dominate an integrated spectrum or spectral energy distribution (SED), the variations in integrated light of a cluster will be dominated by the brightest, most highly-variable members, but will also have significant contributions from fainter and/or lower amplitude stars. 
As of yet, there is no standard treatment of integrated light photometry in variability studies. In this work, we develop a pipeline to extract integrated light curves from star clusters in the Milky Way and the Magellanic Clouds, and present a catalog of star cluster integrated light curves using TESS data. By performing this analysis where resolved photometry from other surveys is available, we demonstrate how the variability of the integrated, unresolved TESS light curves correlates with known individual stellar variability from other observations.

We structure the paper as follows. First we present the data we use in this analysis in Section~\ref{sec:data}, and then describe the process of extracting and assessing the integrated light curves in Section~\ref{sec:light_curves}. We describe our catalog in Section~\ref{sec:cat}, followed by a discussion of the information recovered in the light curves in Section~\ref{sec:info_in_lc}.

\section{Data}
\label{sec:data}
\subsection{Star Cluster Selection}
\label{sec:cluster_selection}
\begin{figure*}[ht]
    \centering
    \includegraphics[width=0.99\textwidth]{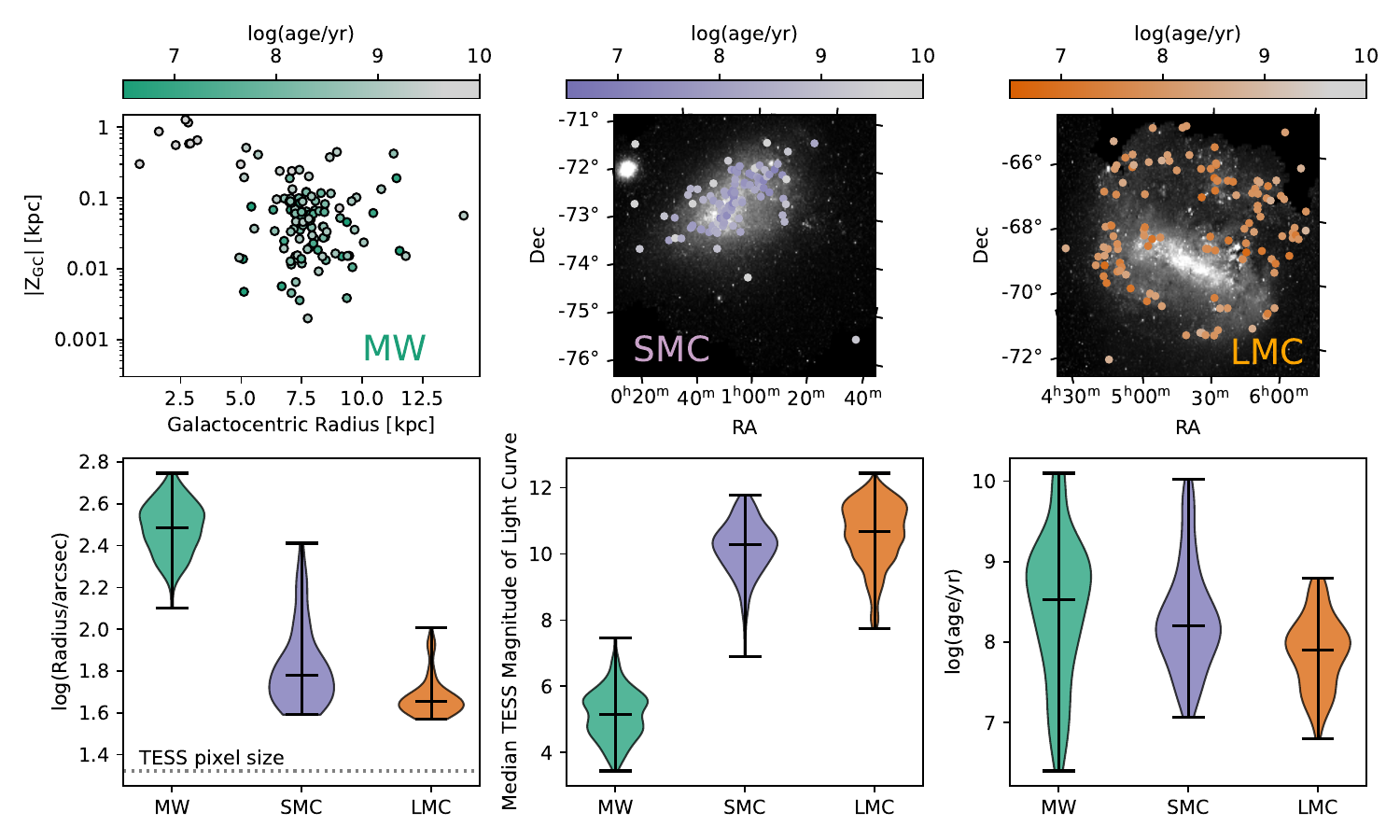}
    \caption{Properties of the clusters in this paper's catalog. The top row shows the distribution of the MW clusters (Section~\ref{sec:mw_clusters}) in Galactocentric radius and height above the plane $|Z_{\rm GC}|$, and of the SMC and LMC clusters (Sections~\ref{sec:lmc_clusters}--\ref{sec:smc_clusters}) in RA and Dec, colored by their ages. The bottom panels show the distributions of cluster angular radius (left), median TESS magnitude of the integrated light (center), and log(age) of the clusters (right).
    Background ($g$-band) images of the LMC and SMC are from the SMASH survey \citep{Nidever_2017_smash,Nidever_2021_smash2}.}
    \label{fig:catalog_clusters}
\end{figure*}

We select the clusters for our light curve catalog from existing catalogs of Milky Way and Magellanic Clouds stellar clusters. These local environments serve as useful laboratories to explore the information available in integrated light curves because we have information on the {\it resolved} stellar populations from other studies. Figure~\ref{fig:catalog_clusters} shows the properties of clusters in our final light curve catalog, based on the selection criteria described here and the light curve quality checks in Section~\ref{sec:lcquality}.

We select Milky Way star clusters from the \citet{kharchenko_global_2013} catalog, where stars were determined to be cluster members by evaluating a membership probability from kinematic, photometric, and spatial criteria, while cluster ages were derived from isochrone fitting using the techniques described in \citet{kharchenko_astrophysical_2005}. Three angular radii were defined for each cluster based on the radial density profile of the member stars. In our cluster selection steps described in Section~\ref{sec:mw_clusters} below, we use both the ``Central Radius'' ($R_{\rm Central}$ here, $r_1$ in the catalog), where the decrease of stellar density flattens, and the ``Cluster Radius'' ($R_{\rm Cluster}$ here, $r_2$ in the catalog), where the surface density of stars becomes equal to the average density of the surrounding field \citep{kharchenko_global_2012}. 

For clusters in the Small Magellanic Cloud (SMC), we start with the catalog of \citet{bica_updated_2020}, which compiles literature values for cluster properties. In the Large Magellanic Cloud, we adopt the \citet{glatt_ages_2010} catalog, whose cluster properties were derived through fitting both Padova \citep{girardi_age_1995} and Geneva \citep{lejeune_database_2001} isochrones to CMDs of the resolved stars. 

Our integrated light curves include flux contributions from everything in the aperture. This includes cluster members, non-member field stars, and background galaxies. In this initial study of integrated light curves, our goal is to select clusters that minimize potential source contamination. Our criteria, detailed below, have the effect of prioritizing clusters identified in the literature that are large and/or bright.

\noindent{\em \textbf{Radius:}} Because of the TESS pixel and image cutout sizes (\ref{sec:lce}), we impose angular radius constraints in our selection across all galaxies. In order to have the full cluster radius and suitable non-cluster regions contained in a single cutout (described in Section~\ref{sec:lce}), we limit the maximum radius of our clusters to $r < 0.25^\circ$. This limit has the added benefit of reducing the non-member contamination. At the same time, our current methodology requires the cluster be larger than a single TESS pixel (21 arcseconds), corresponding to a minimum radius of $r > 0.01^\circ$. In practice, the maximum radius limit only affects the sample of Milky Way clusters (Section~\ref{sec:mw_clusters}), and the minimum radius limit only affects the sample of Magellanic Cloud clusters (Sections~\ref{sec:smc_clusters}--\ref{sec:lmc_clusters}).

\noindent{\em \textbf{Age:}} We require each cluster that we select to have an age estimate in the literature. While this does not disqualify any MW or SMC clusters in our starting catalogs, it does eliminate some LMC objects from \citet{glatt_ages_2010}, where the ages rely on main sequence turnoff fitting and are thus limiting to clusters younger than $\sim$1~Gyr.

Due to the differences in the assembly and contents of the different base cluster catalogs, we cannot establish a truly homogeneous cluster selection process. However, we strive to use broadly similar criteria for each galaxy, outlined below. 

\subsubsection{Milky Way}
\label{sec:mw_clusters}
We start by limiting the \citet{kharchenko_global_2013} sample to those clusters classified therein as ``real star clusters'' or ``globulars''.
We also require at least 100 member stars contained within the $R_{\rm Cluster}$ (i.e., $N_{\rm stars}(<R_{\rm Cluster}) \ge 100$).

We select clusters that are densely concentrated, to minimize the number of field stars and background sources inside our aperture, by comparing the number of stars within $R_{\rm Cluster}$ and within $R_{\rm Central}$: 
\begin{equation}
    \delta = \frac{N_{Cluster} - N_{Central}}{R_{Cluster} - R_{Central}}
\end{equation}

$\delta$ is a rough proxy for the change in the number of stars with angular radius in the cluster, and how similar the number density of stars in the cluster is to the surrounding field region. We visually evaluate the appearance of the clusters for different $\delta$ values, and select a minimum threshold of 2000 for a cluster to be included in our catalog. In short, this process selects clusters that are visually the most easily identified against field stars.

These selection cuts yield a sample of 151 clusters, of which 139 have been observed by TESS. 
The distribution of ages, radii and TESS magnitudes for the Milky Way clusters is shown in green in Figure~\ref{fig:catalog_clusters}.
Stellar densities in the centers of these clusters are typically $\sim$10 stars per TESS pixel (with $G$ < 20).

\subsubsection{Small Magellanic Cloud}
\label{sec:smc_clusters}
We select clusters classified as ``resolved star clusters'' in the \citet{bica_updated_2020} catalog. Furthermore, in order to be consistent in our selection process between galaxies, we set a minimum for the amount of stars. In order to obtain a star count for the SMC clusters, we cross reference stars from \citet{collaboration_gaia_2021} within each cluster radius and select clusters with star counts greater than 300. This produces a sample of 120 clusters. 
The distribution of ages, radii and TESS magnitudes for the SMC clusters is shown in purple in Figure~\ref{fig:catalog_clusters}. 

\subsubsection{Large Magellanic Cloud}
\label{sec:lmc_clusters}
In order to calculate cluster star counts for the LMC clusters, we use the same technique as we did in the SMC. Due to the LMC being $\sim$10~kpc closer than the SMC, we select clusters with a star count greater than 200, as opposed to the 300 used in the SMC. Through visual inspection, we find these two cutoffs to be roughly equivalent in terms of the qualitative detectability of the clusters when examining TESS images. In the LMC, we have a sample of 118 clusters. 
The distribution of ages, radii and TESS magnitudes for the LMC clusters is shown in orange in Figure~\ref{fig:catalog_clusters}.

\subsection{TESS data}
\label{sec:tessdata}
The Transiting Exoplanet Survey Satellite (TESS) is a nearly all-sky photometric survey with the primary objective of discovering transiting exoplanets \citep{ricker_transiting_2015}. TESS observations comprise roughly 27 days of near-continuous measurements, pointing at a $24^\circ \times 96^\circ$ field of view called a ``sector''. TESS has four cameras, each with a field of view of $24^\circ \times 24^\circ$. Each camera has four CCDs, each of which generate $2048 \times 2048$ pixel images with a pixel size of 21$^{\prime\prime}$. For the primary TESS mission (Sectors 1--26), full frame images (FFIs) comprise 30 minute exposures, while for the first extended mission (Sectors 27-55) the cadence was reduced to 10 minutes, and beginning in Cycle 5 (Sectors 56+) the cadence was reduced to 200 seconds. Photometric data are available for objects brighter than $\sim$17th magnitude in the TESS filter, a broad passband ($600-1000$~nm) centered on the Cousins $I$-band filter.

The primary mission of TESS was observing nearly 200,000 unblended stars from the TESS Input Catalog \citep[TIC;][]{stassun_tess_2018}, selected for the purpose of detecting small transiting planets. The FFIs contain tens of millions of detectable stars, although due to the large TESS pixels, many of those stars are blended. There have been many studies focused on blended star extraction \citep[e.g,][]{oelkers_precision_2018,nardiello_psf-based_2020,higgins_localizing_2022}. These efforts have resulted in extracted light curves of more than 20 million FFI sources with relative photometry down to $1\%$ photometric precision  \citep{ricker_transiting_2015, huang_photometry_2020, kunimoto_quick-look_2021}. 

However, despite the success of de-blending techniques, there still exist many sources or stellar populations that are too blended for the above tools to be applied. To the best of our knowledge, TESS data have yet to be used for truly integrated light measurements. In this work, we use the 30~minute cadence FFIs for Sectors 1--26, and the 10 minute cadence FFIs for Sectors 27--39, processed by the TESS Science Processing Operations Center (SPOC) FFI pipeline\footnote{\url{https://heasarc.gsfc.nasa.gov/docs/tess/documentation.html}}. The SPOC FFI pipeline removes systematics due to cosmic rays and CCD effects, and it provides supplemental products for additional systematic correction such as co-trending basis vectors \citep[CBVs, see Section~\ref{sec:lce};][]{jenkins_TESS_2016}. In Section~\ref{sec:light_curves} below, we describe our procedures for extracting light curves for blended ensembles of stars.

\section{Cluster Light Curves}
\label{sec:light_curves}

\subsection{Light Curve Extraction}
\label{sec:lce}
To extract integrated light curves, a number of additional factors must be considered when compared to single-star analyses. We perform our analysis in the form of aperture photometry. The choice of representative ``background'' pixels, and the treatment of spatially- and temporally-varying scattered light when the target and background pixels span a large area on the sky, have a significant impact on the final light curve.  

\begin{figure*}[!ht]
    \centering
    \includegraphics[width=.95\textwidth]{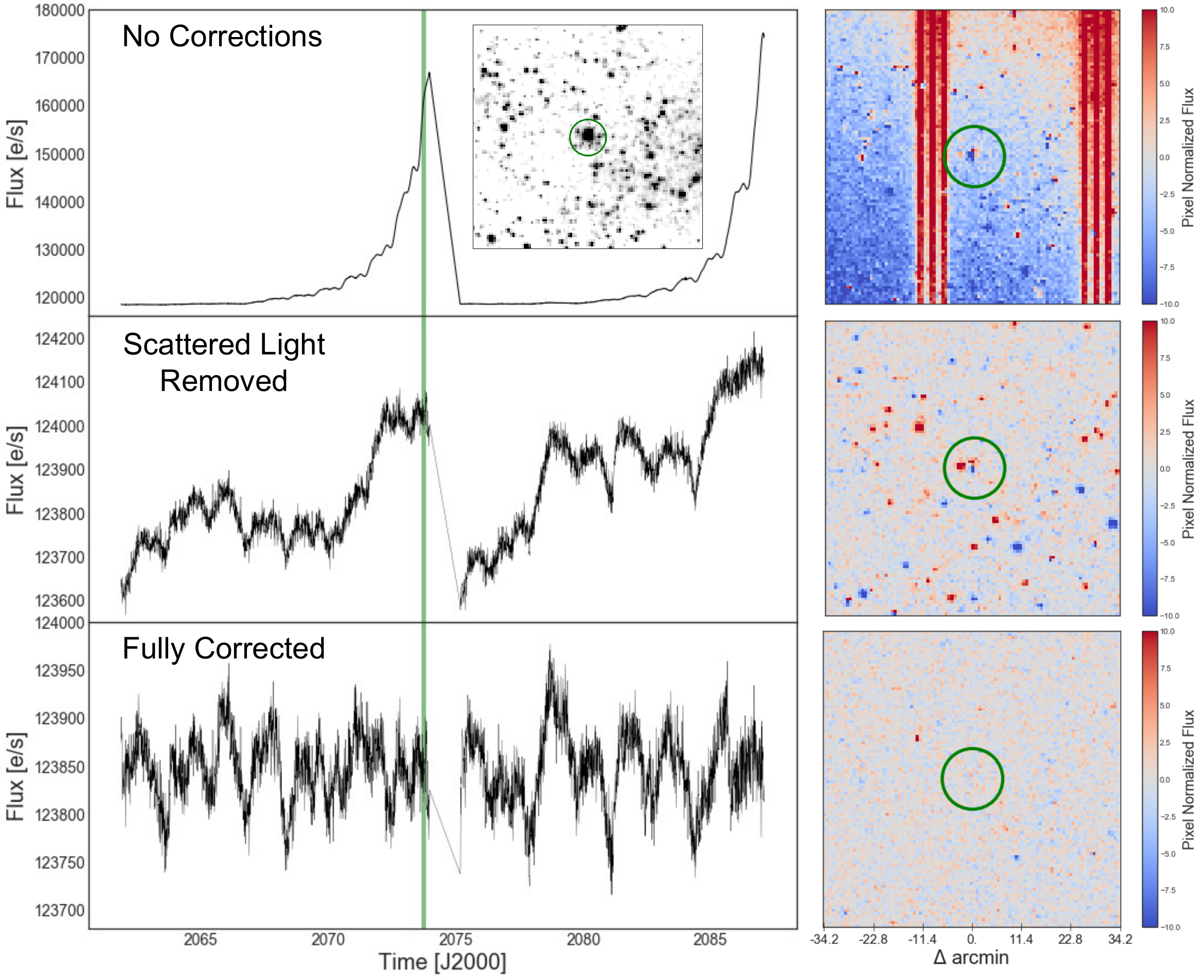}
    \caption{Demonstration of the light curve corrections (Section~\ref{sec:lce}) for the cluster NGC~330. The left column shows light curves for pixels inside the cluster radius (the green circle in the right column). The right column shows normalized individual pixel light curves at the 1700th time step (the green vertical line in the left panels). Inset into the top middle panel is a flux map at this time step of the region, to show the brightness of the cluster compared to the surrounding field. The first row are the raw data, while the second row is the light curve after using PCA to correct for background/scattered light and applying the spline correction. The third row shows the final light curve, with correction from all three elements included. }
    \label{fig:correction}
\end{figure*}

We download the FFI data from MAST\footnote{\dataset[10.17909/
]{\doi{ 10.17909/3y7c-wa45}}} using the \textsc{TESScut} \citep{brasseur_astrocut_2019} feature of the python package \textsc{lightkurve} \citep{lightkurve_collaboration_lightkurve_2018}.  First, for each cluster we extract a $99 \times 99$ pixel cutout (the maximum allowed by MAST’s \textsc{TESScut}) from the FFIs around the cluster central coordinates. We then define an aperture based on the cluster radius from our base catalog\footnote{We emphasize that the definition of ``cluster radius'' is not consistent across cluster studies, or even between our base catalogs. Here we simply adopt the literature values acknowledging their inhomogeneity. However, we experimented with different apertures and note that slightly changing these radii does not significantly change the resulting light curves.}.
An example of this ``raw'' light curve is shown in the top panel of Figure~\ref{fig:correction}.

We use pixels outside of the cluster radius to estimate the background. Specifically, we select the pixels where the median flux is less than the 80th percentile of the entire out-of-aperture pixel sample. This ensures that bright pixels due to, e.g., field stars do not inform our background model.

Due to the spacecraft's $\sim$14 day orbit, the dominant systematic in TESS photometry is spatially-variant transient scattered light from the Earth and Moon, and many efforts have been taken to address this issue in the FFI data \citep[e.g.,][]{hattori_unpopular_2022}. 
We characterize and remove scattered light using the \textsc{lightkurve} package’s RegressionCorrector submodule. In this method, we build a model with the following elements: i) the principal components of the time-series of the background pixels, ii) the TESS mission cotrending basis vectors (CBVs) provided by the SPOC pipeline \citep{jenkins_TESS_2016}, which capture any variability that is common across the CCD, and iii) a model for astrophysical variability comprising a basis-spline component, which flexibly fits for (unspecified) astrophysical variations on a set timescale of less than 10 days. This ``astrophysical'' model is included to enable the systematics vectors (PCA and CBVs) to fit ``around'' any long-term astrophysical variability that there may be in the cluster.

Due to the TESS spacecraft's $\sim$14 day orbit, systematics in TESS data can frequently have periodic signals of 14 days embedded into them. Thus, extracted periods of greater than $\sim$10 days from TESS data can often be unreliable, and so in this work we focus only on short period variability.

These three elements are used to fit the time-series flux for each pixel in the dataset (99 x 99 pixels). The best fit of the systematic (PCA and CBV) components are then used to remove those components only from the dataset (i.e., the spline component is not removed). This then leaves the corrected astrophysical signal, with common systematics across the dataset removed. This step in the correction is shown in the second and third rows of Figure~\ref{fig:correction}.

The number of principal components is a balance --- too few and the systematic trends will not be fully captured, or too many and the data will be overfitted and true astrophysics will be removed. To determine a good balance, we investigate the optimum number of PCA components for several clusters spanning a range of background stellar densities and intensity of scattered light. We studied the cumulative Eigenvalues and found that six principal components could account for 99\% of the background variance in this test suite, even with significant scattered light, and adopt this number for all clusters. 

In order to prevent overfitting, we set priors to our PCA model. Using the \textsc{lightkurve} package, we set priors for mean (\texttt{mu}) and standard deviation (\texttt{sigma}) of the coefficients associated with each linear regression of the principal component design matrix to be the mean and $16{\rm th}-84{\rm th}$ percentile, respectively, of the uncorrected flux distribution. 

These correction processes are demonstrated in Figure~\ref{fig:correction}, where the left hand panels are single-sector light curves representing different steps in the correction. The right hand panels are pixel maps colored by the pixel brightness at the time step highlighted by the green vertical line in the left hand panels, normalized by each pixel's median flux across the entire time span. The first row is the uncorrected light curve, which contains significant scattered light (every $\sim14$ days due to the space craft orbit), as well as significant spacecraft systematics visible as vertical stripes in the right-hand flux map. The second row incorporates the PCA-based correction to subtract the background pixels and the spline corrector, while the bottom row is the final light curve that includes correction from the all of the elements. 

Because this method fits each pixel time-series individually, we are increasing the noise in each pixel. We anticipate this method can be improved by modeling the image series, not individual pixel time-series. However, for our purposes, we find this method a) adequately removes the spacecraft systematics and b) adequately preserves the stellar signal (see Figure~\ref{fig:correction} and Section~\ref{sec:lc_stats}). 

\subsection{\texttt{elk}: Publicly Available Python Package for Integrated Light Curve Extraction and Diagnosis}
\label{sec:elk}
The workflow described in Section~\ref{sec:lce} is made publicly available as a Python package: int\textbf{E}grated \textbf{L}ight \textbf{K}urve, or \texttt{elk}.

This package can be used to download and correct the integrated TESS light curve for any given aperture smaller than 0.57~degrees in diameter, the maximum allowed cutout size of FFI data using the \textsc{TESScut} tool. In this work, we have chosen to focus on the study of star clusters, but \texttt{elk} can be used for any study of integrated light from an astronomical population. We include options for altering a variety of settings including: the aperture radius, the TESS cutout size, the number of PCA components to use in the design matrix, whether to use the spline corrector, and how many knots to include in the spline corrector. One can additionally use \texttt{elk} to visualize light curves, as well as their auto-correlation functions and the Lomb-Scargle periodogram \citep[LSP;][]{lomb_least-squares_1976, scargle_studies_1982}, the primary analysis tool we use to identify periodic variability (see more discussion of the LSP in Section \ref{sec:lsp_maps}). The package also includes functionality for computing a series of commonly used variability statistics, as well as novel techniques for diagnosing integrated light curves (Section~\ref{sec:lsp_maps}).

Full documentation for the package, with tutorials and a comprehensive user guide, is available online\footnote{\url{https://elk.readthedocs.io/}}. The package can be installed via \texttt{pip install astro-elk} and the source code is accessible on GitHub\footnote{\url{https://github.com/tobin-wainer/elk}}. \texttt{elk} is jointly published in JOSS (Wainer et al. 2023, in preparation).

\subsection{Quality Assurance}
\label{sec:lcquality}
Within the sample of 389 clusters selected in Section~\ref{sec:cluster_selection}, each cluster was observed by TESS for a median of four sectors. Each time a cluster is observed, we define one `sector of observation'. Clusters in the LMC, which is located in the southern TESS continuous viewing zone (S-CVZ), were typically observed over 25 sectors, while clusters in the Milky Way were observed for an median of three sectors. We processed each sector of observation of a given cluster independently, for a total of 3436 unique cluster sectors extracted. Due to the methodology of extracting light curves from integrated light sources combining many TESS pixels, the cluster light curves can be more susceptible to systematics compared to a single source. To ensure the quality of the resulting light curves, we remove observations which contain a large amount of systematic trends, as determined from the following tests. 

First, if a cluster is located too close to the edge of a TESS detector, we cannot perform a precise and uniform background subtraction as we do not have a full field of background pixels. Therefore, we remove any observations where the target cluster is within 0.5~degrees of the edge of a detector. Furthermore, through visual inspection of all light curves, we found significant systematics present in all Sector 1 observations that we were unable to remove with the procedure described in Section~\ref{sec:lce}, and remove these from our analysis. 

We also check for sectors with significant scattered light (generally from Earth or the Moon) that is not cleaned by the procedures in Section~\ref{sec:lce}. If there is uncorrected scattered light, there will be spatially dependent flux differences across the cutout (e.g., top right panel of Figure~\ref{fig:correction}). Therefore, to test for this difference, we fit a 2D model to the background pixels in our cutout at each time step, defined by
\begin{equation}
    Z = C_1 \times X + C_2\times Y + C_3
\end{equation}
where $X$ and $Y$ are the row and column numbers of the pixel location we are fitting, and $Z$ is the flux at that time step scaled by each pixel's maximum value (as in Section~\ref{sec:lce} and the flux maps in Figure~\ref{fig:correction}). Through visual inspection, we define a cutout to have significant un-removed scattered light if both $C_1$ and $C_2$ are greater than $0.02$, or if $C_3$ is greater than $2.5$. We remove a total of 712 sector observations due to the presence of this scattered light. 

There are a total of 29 clusters where none of the observations pass these quality checks: 15 in the Milky Way, and 14 in the SMC. Roughly 60\% of these clusters were removed for not having a sector pass the scattered light test, while the remaining 40\% were for failing a combination of quality tests. Our final catalog has at least one sector light curve for 348 clusters (Figure~\ref{fig:catalog_clusters}).

\section{Light Curve Catalog}
\label{sec:cat}

Using the techniques described in Section~\ref{sec:light_curves}, we present sector light curves for 124 Milky Way clusters, 106 SMC clusters, and 118 LMC clusters. We detail which sectors the light curves are from, as well as the light curve time span. Due to the position of the LMC in the S-CVZ, each LMC cluster has nearly continuous coverage from TESS, yielding many more observations compared to the MW and SMC clusters. We also provide the standard deviation and $\rm 5th - 95th$ percentile range of the normalized light curves, as proxies for the variability in the flux distributions. 
These tables are previewed in Tables~\ref{tab:mw_cat}--\ref{tab:lmc_cat}.

\begin{deluxetable*}{ccccccccccc}
\tabletypesize{\scriptsize}
\setlength{\tabcolsep}{0.05in}
\tablewidth{0pt}
\tablecaption{Milky Way Cluster Light Curve Catalog \label{tab:mw_cat}}
\tablehead{
\multirow{2}{*}{Name} & \colhead{RA\tablenotemark{a}} & \colhead{DEC\tablenotemark{a}} & \colhead{Radius\tablenotemark{a}} & \colhead{Age\tablenotemark{a}} & \colhead{Obs} & \colhead{$\#$ Good} & \multirow{2}{*}{Good Sector Observations\tablenotemark{b}} & \colhead{Light Curve Lengths} & \multirow{2}{*}{Normalized Stdev} & \multirow{2}{*}{Normalized Range} \\
 & [deg] & [deg] & [deg] & [log(yr)] & Avail & Obs & & [Num timesteps] & &  
} 
\startdata
ASCC 116 & 329.625 & 54.49 & 0.14 & 7.95 & 4 & 4 & [TMS-16, TMS-17, TMS-56...] & [935, 1126, 11766...] & [0.000621, 0.000704, 0.00068...] & [0.002056, 0.002352, 0.002213...] \\
ASCC 57 & 152.73 & -66.7 & 0.215 & 9.26 & 6 & 3 & [TMS-11, TMS-36, TMS-38] & [1180, 3467, 3693] & [0.000182, 0.000218, 0.000216] & [0.000598, 0.000719, 0.000713] \\
ASCC 8 & 35.19 & 59.69 & 0.24 & 7.77 & 2 & 2 & [TMS-18, TMS-58] & [1103, 11692] & [0.000214, 0.000162] & [0.00057, 0.000534] \\
ASCC 81 & 236.692 & -50.995 & 0.16 & 8.75 & 2 & 2 & [TMS-12, TMS-39] & [1234, 3865] & [0.000373, 0.000649] & [0.001216, 0.002036] \\
ASCC 9 & 41.745 & 57.745 & 0.21 & 7.0 & 2 & 1 & [TMS-18] & [1103] & [0.000906] & [0.002909] \\
\enddata

\tablecomments{Table \ref{tab:mw_cat} is published in its entirety in the electronic edition of the {\it Astrophysical Journal}. A portion is shown here for guidance regarding its form and content. The ellipses indicate more data available, not shown here.}
\tablenotetext{a}{Cluster parameters are taken from the \citet{kharchenko_global_2013} catalog.}
\tablenotetext{b}{TMS refers to TESS Mission Sector as it appears in MAST.}
\end{deluxetable*}

\begin{deluxetable*}{ccccccccccc}
\tabletypesize{\scriptsize}
\setlength{\tabcolsep}{0.05in}
\tablewidth{0pt}
\tablecaption{Small Magellanic Cloud Light Curve Catalog \label{tab:smc_cat}}
\tablehead{ 
\multirow{2}{*}{Name} & \colhead{RA\tablenotemark{a}} & \colhead{DEC\tablenotemark{a}} & \colhead{Radius\tablenotemark{a}} & \colhead{Age\tablenotemark{a}} & \colhead{Obs} & \colhead{$\#$ Good} & \multirow{2}{*}{Good Sector Observations\tablenotemark{b}} & \colhead{Light Curve Lengths} & \multirow{2}{*}{Normalized Stdev} & \multirow{2}{*}{Normalized Range} \\
 & [deg] & [deg] & [deg] & [log(yr)] & Avail & Obs & & [Num timesteps] & &  
} 

\startdata
OGLE-CL SMC  324 & 6.18083 & -73.75389 & 0.03 & 9.15 & 4 & 2 & [TMS-27, TMS-28] & [3351, 3449] & [0.000481, 0.000508] & [0.001557, 0.00158] \\
OGLE-CL SMC  319 & 6.19167 & -72.79389 & 0.06 & 9.81 & 3 & 2 & [TMS-27, TMS-28] & [3351, 3449] & [0.000253, 0.000276] & [0.000814, 0.000901] \\
OGLE-CL SMC  311 & 6.70208 & -71.53472 & 0.06 & 10.02 & 3 & 2 & [TMS-27, TMS-28] & [3351, 3449] & [0.000287, 0.000239] & [0.000906, 0.000774] \\
NGC152 & 8.23458 & -73.11583 & 0.05 & 9.09 & 3 & 2 & [TMS-27, TMS-28] & [3351, 3449] & [0.000301, 0.000264] & [0.000978, 0.000866] \\
HW8 & 8.445 & -73.63278 & 0.03 & 8.0 & 4 & 2 & [TMS-27, TMS-28] & [3351, 3449] & [0.000662, 0.000637] & [0.002163, 0.002117]
\enddata

\tablecomments{Table \ref{tab:smc_cat} is published in its entirety in the electronic edition of the {\it Astrophysical Journal}.  A portion is shown here for guidance regarding its form and content.}
\tablenotetext{a}{Cluster parameters are taken from the \citet{bica_updated_2020} catalog.}
\tablenotetext{b}{TMS refers to TESS Mission Sector as it appears in MAST.}
\end{deluxetable*}

\begin{deluxetable*}{ccccccccccc}
\tabletypesize{\scriptsize}
\setlength{\tabcolsep}{0.05in}
\tablewidth{0pt}
\tablecaption{Large Magellanic Cloud Cluster Light Curve Catalog \label{tab:lmc_cat}}
\tablehead{
\multirow{2}{*}{Name} & \colhead{RA\tablenotemark{a}} & \colhead{DEC\tablenotemark{a}} & \colhead{Radius\tablenotemark{a}} & \colhead{Age\tablenotemark{a}} & \colhead{Obs} & \colhead{$\#$ Good} & \multirow{2}{*}{Good Sector Observations\tablenotemark{b}} & \colhead{Light Curve Lengths} & \multirow{2}{*}{Normalized Stdev} & \multirow{2}{*}{Normalized Range} \\
 & [deg] & [deg] & [deg] & [log(yr)] & Avail & Obs & & [Num timesteps] & &  
} 

\startdata
NGC1652 & 69.59167 & -68.6725 & 0.01 & 8.5 & 26 & 15 & [TMS-2, TMS-3, TMS-9...] & [1196, 1077, 1084...] & [0.000515, 0.000545, 0.000544...] & [0.001688, 0.001725, 0.001736...] \\
SL63-37 & 71.69167 & -72.39472 & 0.02 & 8.7 & 28 & 17 & [TMS-2, TMS-3, TMS-7...] & [1196, 1077, 1086...] & [0.000382, 0.000457, 0.001728...] & [0.001265, 0.001408, 0.003991...] \\
NGC1695 & 71.93333 & -69.37389 & 0.01 & 8.0 & 30 & 17 & [TMS-2, TMS-3, TMS-4...] & [1196, 1077, 1027... ] & [0.000492, 0.000646, 0.000556...] & [0.001669, 0.00171, 0.001799...] \\
NGC1698 & 72.26667 & -69.11472 & 0.01 & 8.0 & 29 & 23 & [TMS-2, TMS-3, TMS-4...] & [1196, 1077, 1027...] & [0.001776, 0.004937, 0.004642...] & [0.005842, 0.017143, 0.017566...] \\
NGC1704 & 72.47917 & -69.75528 & 0.01 & 7.5 & 29 & 18 & [TMS-3, TMS-5, TMS-6...] & [1077, 1176, 980...] & [0.002961, 0.00437, 0.002454...] & [0.005152, 0.01412, 0.006831] \\
\enddata

\tablecomments{Table \ref{tab:lmc_cat} is published in its entirety in the electronic edition of the {\it Astrophysical Journal}. A portion is shown here for guidance regarding its form and content. The ellipses indicate more data available, not shown here.}
\tablenotetext{a}{Cluster parameters are taken from the \citet{glatt_ages_2010} catalog.}
\tablenotetext{b}{TMS refers to TESS Mission Sector as it appears in MAST.}
\end{deluxetable*}

\subsection{Light Curve Flux Distribution Statistics}
\label{sec:lc_stats}

Many studies have adopted the standard deviation and the flux range of a light curve to determine the level to which the light curve shows variability \citep[e.g.,][]{mcquillan_statistics_2012, sokolovsky_comparative_2017}. 
For each sector light curve in our catalog, we normalize by the median flux and calculate the standard deviation and $\rm 5th-95th$ percentile range (hereafter referred to as the variability ``metrics''). This normalization removes the fractional flux variation due to distance differences. In this section, we will inspect these metrics, and their sensitivity to systematics. We note that there are {\it many} more ways to quantify the variability of a source \citep{sokolovsky_comparative_2017}, and a full characterization of variability of each cluster will be the focus of future studies; for this initial study, we simply want to understand the reliability of the light curves.

\begin{figure}
    \centering
    \includegraphics[width=0.45\textwidth]{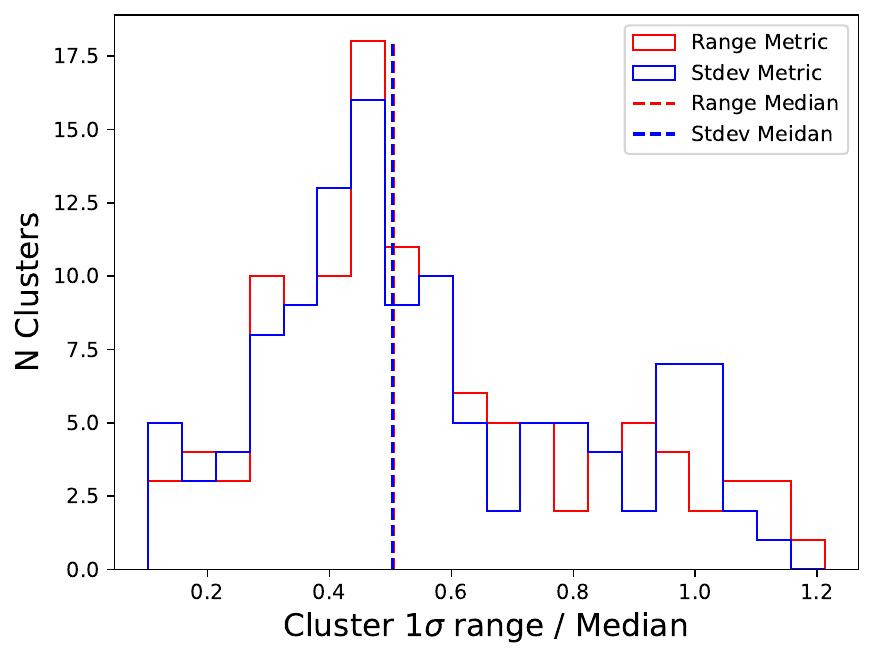}
    \caption{
    Distribution of the 1$\sigma$ uncertainty for LMC cluster range and standard deviation metrics (Section~\ref{sec:lc_stats}), demonstrating the consistency in which the flux distribution metrics are consistent across each available sector of observation.
    } 
    \label{fig:dist_of_dists}
\end{figure}

Due to the unique location of the LMC in the S-CVZ, nearly every LMC cluster has more than 20 sectors of observation per cluster. We took advantage of this large number of sectors to explore the consistency of the metrics, and obtained an uncertainty by calculating the 1~$\sigma$ spread in metric values across individual sectors for each cluster. We then scale the 1~$\sigma$ range by the metric median to provide the relative difference for each cluster. The distribution for all LMC clusters is shown in Figure~\ref{fig:dist_of_dists}. In blue is the scaled 1~$\sigma$ variation of flux standard deviation, measured across all sectors for a given cluster (median relative difference of 0.50), and in red is the same quantity for the $\rm 5th-95th$ percentile range (median relative difference of 0.50). 

We explored the impact that light curve time duration has on the metrics. To do this comparison, for a small number of clusters we take a full sector light curve (with roughly 3000 time steps), and divide it into four sub-sections of 6.75~days each, which is shorter than our shortest light curves ($\sim$700 time steps),
We then calculate the metrics for each sub-section separately. We find the metrics to be consistent with each other (within 1\%) across all sub-sections of a single sector light curve, with differences smaller than the scatter between different sectors described in the previous paragraph. We take this consistency to indicate that these metrics, at least, are not dominated by counting noise in even our shortest sectors.

Further, we compared the signal-to-noise ratio (SNR) of the sector observation to the metrics. For each time step in an observation, we take the flux divided by the flux error (after corrections; Section~\ref{sec:lce}) as the measure of SNR and calculate the median SNR for the observation. We find that clusters in the Milky Way have a SNR typically $\sim$10$\times$ higher than the clusters in the SMC/LMC. Further, there is a correlation between the variability metrics and sector observation SNR for some of the clusters in the LMC and SMC with very low SNR. This is not unexpected, as the Gaussian noise of a distribution will dominate the standard deviation for any uncertainty greater than the underlying variability. 

\begin{figure}
    \centering
    \includegraphics[width=.47\textwidth]{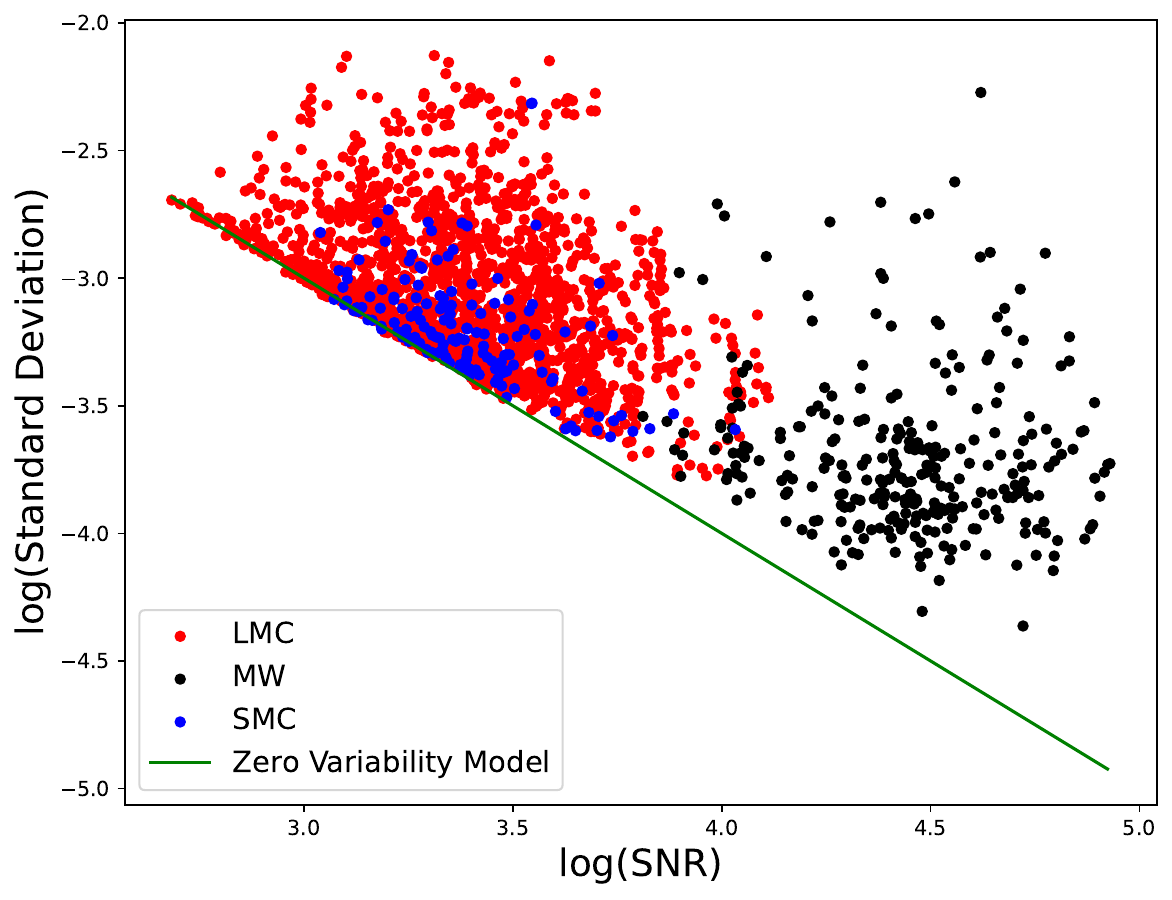}
    \caption{The normalized light curve standard deviation as a function of $\log{(\rm SNR)}$ (Section~\ref{sec:lc_stats}). Each point represents a sector observation: black for the Milky Way clusters (which have the highest SNR due to their close proximity), blue for SMC clusters, and red for LMC clusters. The green line represents the theoretical limit for a light curve being dominated by purely Gaussian noise.  }
    \label{fig:s_n}
\end{figure}

This threshold is represented by the green line in Figure~\ref{fig:s_n}, where we assume a flat light curve with a given Gaussian noise estimate, and the resulting light curve standard deviation. We believe the sharpness of this threshold shows that our SNR measurements are a reasonable estimate of the true SNR. There are a number of clusters that are at this limit, where this metric is only probing the Gaussian noise of the light curve. However, the majority of clusters lie above this limit, implying the standard deviations reflect genuine variability beyond the noise. 

In addition to quantifying the SNR for our cluster sample, we assess the impacts the SNR has on the overall flux distribution. We want to know at what noise level an observed flux distribution is significantly different from the intrinsic distribution. For this test, we create a synthetic cluster light curve (see Section~\ref{sec:low_amp_variability}) and run a bootstrap sampling with decreasing SNR. We perform a Kolmogorov-Smirnov (K-S) test on the resulting bootstrapped light curves compared to the original. Due to inherent randomness in the re-sampling, we perform 100 trials. The median pvalue of these 100 trials is greater than 0.99 from SNR of 10000 to 6.25, where it drops precipitously until reaching 0.1 at SNR equal to 2.

\section{Information Recovered in Integrated Light Curves}
\label{sec:info_in_lc}

One interesting question this light curve catalog will help answer is whether the variability characteristics of integrated light curves can be correlated to physical properties of the underlying stellar population. Because the catalog assembled here is limited to clusters in the Galaxy and the Magellanic Clouds, we have population-level data for these clusters, such as age and metallicity. For clusters in the Milky Way, we also generally have properties of individual stars in these clusters, including surface temperature and evolutionary state, as well as likelihood of membership in the cluster. Therefore, we can investigate how the variability of the integrated, unresolved TESS light curves of the Milky Way clusters correlates with known individual stellar variability from other observations.

\subsection{Spatial Maps of LSP Power}
\label{sec:lsp_maps}

\begin{figure*}[!ht]
    \centering
    \includegraphics[width=\textwidth]{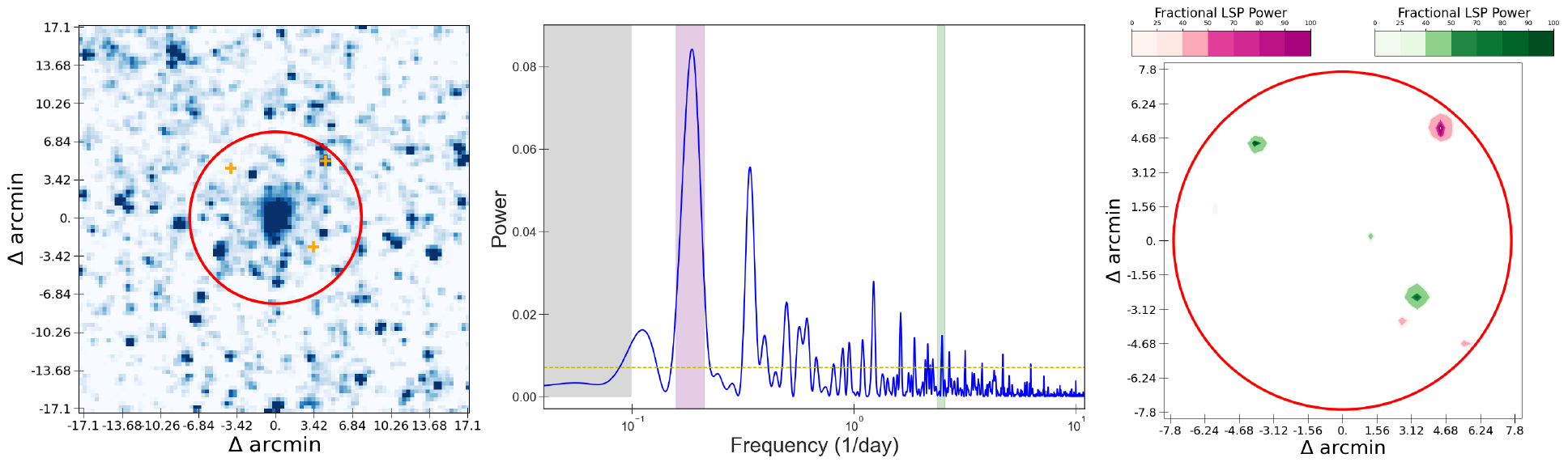}
    \caption{Identification of individual pixel power contributing to frequency ranges in the LSP for NGC 6304. In the left panel is a TESS flux map for NGC 6304 (centered at $(\alpha,\delta)=(258.635, -29.462)$), with the cluster radius of 0.13~degrees shown by the red circle. The orange crosses mark the location of the three largest contours in the right panel. In the middle panel is the integrated light curve's LSP. The dashed horizontal yellow line shows the false alarm probability of 0.007, and the grayed-out region is the part of frequency space ($ >10$ day periods) we are not sensitive to.  Two frequency ranges are highlighted, with 0.183 to 0.199 days$^{-1}$ highlighted in purple and 2.367 to 2.608 days$^{-1}$ highlighted in green. The right-hand panel shows the spatial distribution of power in these frequency ranges, in the corresponding colors. The red circle has the same angular size as in the left panel. The purple and green contours show pixel power from the corresponding frequency ranges in the middle plot.  The feature in purple corresponds to a low mass X-ray binary candidate \citep{guillot_x-ray_2009, heinke_x-ray_2020}, and the second feature corresponds to two RR Lyrae stars whose literature frequencies are close together, but are well-separated in the TESS image \citep{collaboration_gaia_2021}.  See Section~\ref{sec:lsp_maps} for more details.}
    \label{fig:gif_snap}
\end{figure*}

One way to verify the persistence of this variability information is to search for the signature of known cluster variables in the integrated cluster light curves. If the variability signature is still prevalent, we can confirm that we can trace specific signals in the integrated light curves to the astrophysical variability of a single star. We use the Lomb-Scargle periodogram (LSP) to analyze the periodic signals in our integrated light curves, implemented in the \texttt{astropy.timeseries} module \texttt{LombScargle}. We note that the strength of a peak in the LSP of the integrated light curve depends on not only the strength of the variability of the star (amplitude and coherence) but also the relative brightness of the variable star within the aperture.

Our first step is to examine the integrated light curve of a cluster with at least one known variable star and see if there is a peak in the LSP corresponding to the known period of the star(s). We select variable stars for this analysis because of the known periodicity, and large amplitude variation, where the truth values are well documented. We demonstrate our method for the globular cluster NGC~6304 (Figure~\ref{fig:gif_snap}), which has multiple confirmed and characterized variable stars \citep{clement_variable_2001,clement_vizier_2017}. There are two RR~Lyrae (RRL) stars listed in \citet{clement_vizier_2017}\footnote{We have cross-matched both of these stars with Gaia confirmed RRL (Gaia DR3 $4107310867702985600$ and Gaia DR3 $4107362540579204480$) and confirmed their positions with the Gaia coordinates.}, both of which have pulsational frequencies close to 2.56 days$^{-1}$. In the middle panel of Figure~\ref{fig:gif_snap}, we can clearly see a peak in the LSP corresponding to the known frequency of these two RRLs highlighted in green. We note that there are also several other peaks in the LSP, most of which are due to the other $\sim20$ RRLs in the cluster. This experiment confirms that in an integrated light curve, the underlying stellar variability can be detectable. 

One benefit of performing this test in the MW is our ability to know which stars are contained in unresolved TESS pixels. In integrated light curves containing many stars, there are likely multiple stars exhibiting variability on similar time scales. For our example cluster NGC~6304, while we demonstrated that there is a peak in the LSP with the known frequency of the RR~Lyrae, we want to confirm that this signal is coming from the sky location of the two RR~Lyrae, and not simply a coincidence. To do this, we develop a tool to display pixels in the integrated aperture with significant LSP power at specific frequencies (available in \texttt{elk}; Section~\ref{sec:elk}). This can be seen in the right panel of Figure~\ref{fig:gif_snap}, where we highlight the specific pixels contributing to two peaks in the LSP in the middle panel. (This panel is essentially a map of the power in narrow frequency ranges, measured from individual {\it pixel} light curves.) The first peak, shown by the frequency range in purple, is mostly attributed to the pixels in the upper right of the aperture, represented by the purple contours. Within these pixels is a low-mass X-ray binary candidate \citep{guillot_x-ray_2009, heinke_x-ray_2020} \footnote{This low-mass X-ray binary candidate has an associated TIC ID (78302251) and a TESS QLP light curve available in MAST. However, according to LSP analysis the QLP light curve has a periodic signature at 0.11 days$^{-1}$, not at 0.19 days$^{-1}$ as in our integrated light curve.}. 

The second strongest peak (at 0.34 days$^{-1}$, not highlighted in the figure), is associated with an eclipsing binary (BLG-ECL-2201, DR3 4107356115130651904) with a literature frequency of 0.16 days$^{-1}$ \citep{soszynski_ogle_2016}, close to double that of the peak in the LSP. Since eclipsing binaries have two cycles of peaks and troughs in a single orbital period, a L-S analysis will typically detect half the true period \citep{fetherolf_variability_2022}. The peak in the LSP highlighted in green corresponds to two separate green contour regions in the right hand panel. These two locations map to sky positions of the two RR~Lyrae stars \citep{clement_variable_2001, collaboration_gaia_2021, castro-ginard_hunting_2022} discussed above. This tool can therefore match features in the LSP to variable objects, and will be used in the following sections to cross-match high-amplitude variables to features in the integrated light curve's LSP. 

We note also that the two RRL stars indicated in Figure~\ref{fig:gif_snap} are not the bright pixels shown in the left panel to the upper left and lower right of the cluster center, but are offset, with locations indicated by the orange crosses. However, the low mass X-ray binary candidate \citep{guillot_x-ray_2009, heinke_x-ray_2020} is in fact located at the position of the bright pixels to the upper right near the edge of the aperture. This reinforces that significant features in the LSP can arise from sources with a range of brightness, depending on the characteristics of their variability. It is noteworthy that the tool can distinctively pick out the two RRL stars with the same frequency, even though they are blended with tens of other stars, and they are not the brightest stars in the cluster. 

To determine the robustness of the LSP for this example, we calculate the false alarm probability (FAP) for spurious peak strengths \citep{horne_prescription_1986}. We run 100 trials and find that the $1\%$ FAP power for our NGC~6304 light curve is 0.007, and thus both peaks in the LSP we have discussed are statistically significant. To determine the uncertainty in the LSP power, we run a bootstrap sampling for 1000 trials using the flux errors provided in the TESS FFI. We define the $84$th - $16$th percentiles of the trials as our $1\sigma$ uncertainty, and run a frequency range from $0.04-11$~days$^{-1}$, with a step size of $0.001$. Across all time steps, we find the median uncertainty in the LSP power across all frequencies to be 10$^{-3}$ (dimensionless). 

This example demonstrates the ability to map features in the LSP of the integrated light curve to specific stars. However, we observed cases (in this and other clusters) where features cannot be easily attributed to individual stars, but rather, to an area in the cluster encompassing many pixels and many unresolved stars. 
While there are many more ways to characterize the variability of a source, these initial tests demonstrate that phase mixing of multiple sources does not always drown out the signal from individual sources. Analyzing the individual sources of variability for each of these clusters will be a focus of future studies. The remainder of this section will explore the recoverable information from high-amplitude variables, as well as information about what part of the CMD a cluster occupies. 

This tool is made public as a part of the \texttt{elk} package (Section~\ref{sec:elk}), with an automated peak finder, and an option to perform a Simbad query for stars that are contained within the pixels contributing power to LSP peaks. 

\subsection{Cross Match to Known Cluster RR Lyrae}
\label{sec:rr_ls}

\begin{figure*}
    \centering
    \includegraphics[width=.99\textwidth]{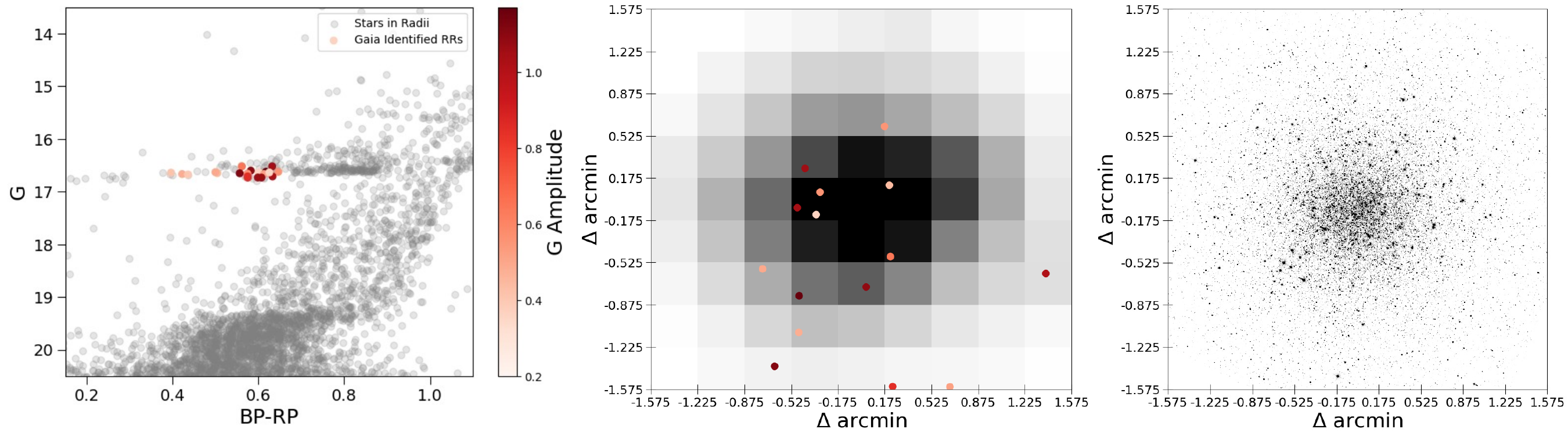}
    \caption{The color magnitude diagram for NGC~1261 is shown on the left. All of the Gaia stars are shown in gray, and the 20 RRL stars discussed in Section~\ref{sec:rr_ls} are colored according to the amplitude of their $G$-band photometric variability. The middle planel is the TESS flux map with points indicating the location of the RRL stars from the left panel. On the right is the HST F606W image, mapped to the sky position of the TESS image. Both images are centered at $(\alpha,\delta)=(48.060, -55.225)$, covering the central $\sim3$ arc-minutes of the cluster.} 
    \label{fig:gaia_cmd}
\end{figure*}

\begin{figure*}[!ht]
    \centering
    \includegraphics[width=\textwidth]{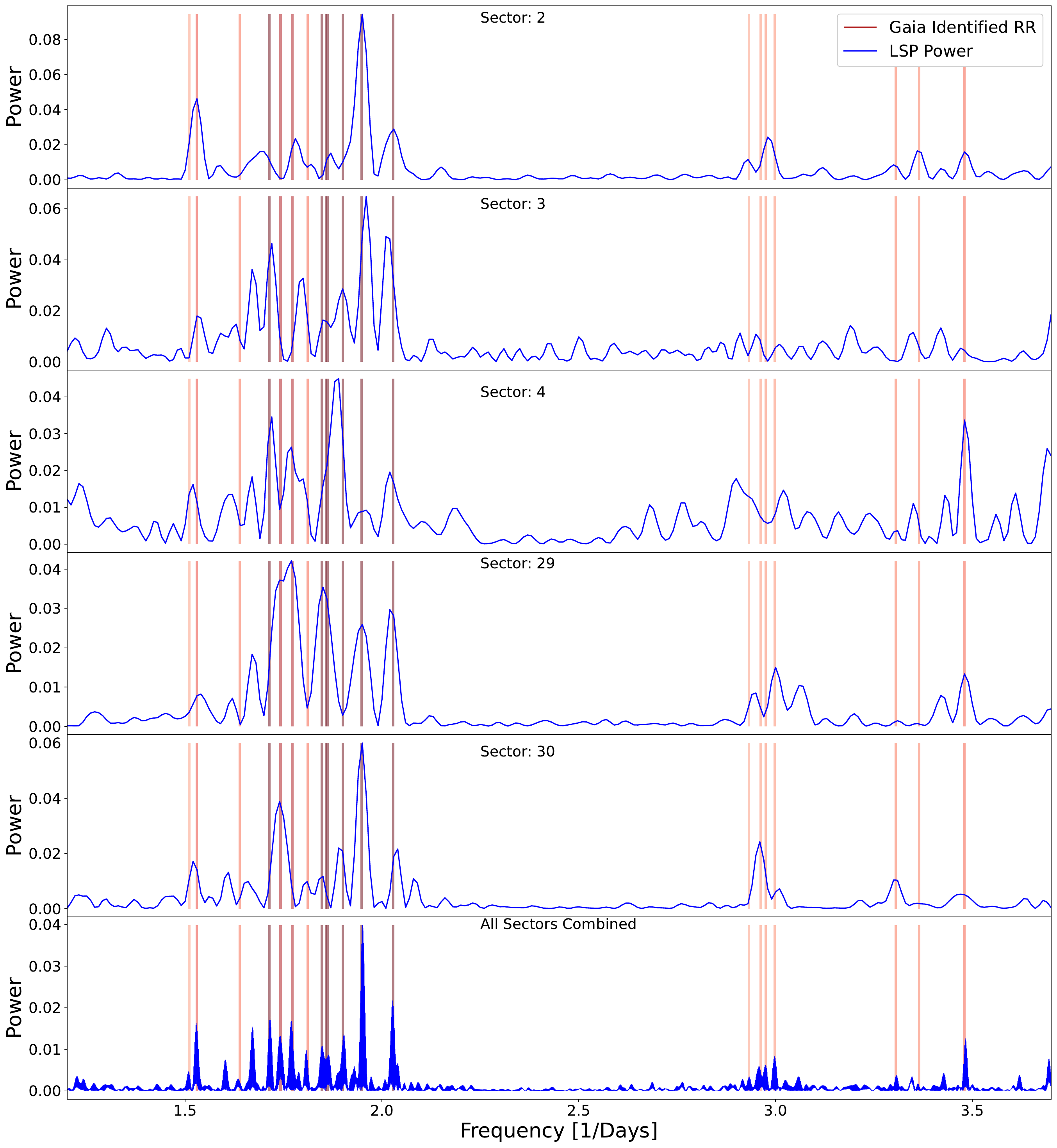}
    \caption{The LSP for each sector of observation for NGC~1261 (Section~\ref{sec:rr_ls}). The vertical lines in each panel correspond to literature frequencies of the cluster's 20 RRL stars, and are colored by the $G$-band amplitude of variation seen in Figure~\ref{fig:gaia_cmd}.  The bottom panel is the LSP for the sector stitched light curve, sampled more finely due to the increased baseline of considering multiple sectors. }
    \label{fig:rrs}
\end{figure*}

In this section, we expand upon the LSP analysis above for an example cluster with many RRL stars. RRL stars are good sources for this analysis because their $\sim$3~day periods mean they complete multiple full cycles in a single TESS sector. Furthermore, because of their lower masses, RRL appear in much older clusters, particularly globular clusters (which are much denser on the sky than open clusters), and are typically not the brightest stars in their cluster. Thus, RRL stars are also a good test of blending and phase mixing in dense crowded regions. 

Here, we analyze the globular cluster NGC~1261. This cluster is not a member of our final light curve catalog because the high density of stars in the globular cluster lead to a mischaracterization of star counts in the \citet{kharchenko_global_2013} catalog. However, NGC~1261  contains a high number of RRL stars with known periods, and there are no other significant variable stars reported in the \citet{clement_variable_2001} or \citet{clement_vizier_2017} catalogs for this cluster, making it an excellent testing ground for RRL stars.
We select the 20 of 22 Gaia-identified RRL stars \citep{eyer_gaia_2022,clementini_gaia_2022} that lie on the horizontal branch\footnote{The other two Gaia-flagged RRL stars are much fainter, with $G \sim 19.5$, and may in fact not be cluster RRL stars.}, which are the largest-amplitude variables in the cluster \citep{clement_variable_2001, clement_vizier_2017}. This cluster CMD is shown in Figure~\ref{fig:gaia_cmd}, where the RRL stars are colored by the G band amplitude of their photometric variation. There are five ``good'' sectors of observation for NGC~1261 (Section~\ref{sec:lcquality}); the LSPs for each of these are shown in Figure~\ref{fig:rrs}, where the dashed vertical lines lie at the Gaia identified frequencies of the RRL stars and are shaded as in Figure~\ref{fig:gaia_cmd}. In each panel of Figure~\ref{fig:rrs}, we can see that the majority of the peaks in the LSP correspond to frequencies of known RRL. The center and right panels of Figure~\ref{fig:gaia_cmd} also highlight that these recovered RRL are distributed throughout the cluster, including in the dense center. 

Each sector observation has a slightly different LSP, and RRL stars with a strong peak in one sector may have a weaker or non-existent peak in another. For example, around a frequency of $\sim3$~days$^{-1}$, there is a clear peak coming from four different RRL stars in all sectors except for Sector 3, in which this feature is indistinguishable from the rest of the LSP. This type of phenomenon is not necessarily unexpected. Each sector has its own patterns of systematic noise and instrumental trends, which will obfuscate specific ranges of periodicity from detection by the LSP. Furthermore, in integrated measurements, phase mixing can reduce the periodogram power of variables with similar periods. That is, at certain epochs the flux coming from individual stellar light curves will be added destructively, and constructively at others, meaning that the $\sim27$ day snapshots taken with TESS are likely to have instances where phase mixing washes out periodic signal. 

The bottom panel of Figure~\ref{fig:rrs} is the LSP for the combined light curve comprising {\it all} of the above sectors, sampled at 0.0001~days$^{-1}$ to account for the longer baseline associated with the stitched light curve. The LSP for the combined light curve has much narrower peaks due to large number of photometric measurements over a long time baseline, and the smoothing of individual sector systematics. For example, there are two RRL stars\footnote{Gaia DR3 4733794859231637376 and 4733794691727354624, respectively} with similar frequencies of 1.51 and 1.53~days$^{-1}$. In each individual sector, there is a broad peak spanning both frequencies, but in the sector-stitched version, there is a double peak, with the literature frequencies for each star being clearly distinguished in the LSP. Further, the RRL star with the larger amplitude of variation has more LSP power than the RRL star with the lower amplitude of variation. Indeed, for most of the RRL stars in the cluster, there is a clear peak in the LSP of the stitched light curve. This stitched light curve is less susceptible to phase mixing issues since the constituent light curves have a wider range of phase offsets during the longer time baseline. 

While there are many stars within the selected aperture, the example of this cluster shows that periodic signals from RRL stars are not generally being washed out by the large number of sources. Even though there are many different periodic sources varying in different phases, the signals from many of the RRL stars are still distinguishable in individual sectors, and especially so in the longer-baseline combined light curve. 

\subsection{Cross Match to Known Cluster Cepheids}
\label{sec:cepheids}
 
Cepheid variable stars are prominent astrophysical tracers due to their young ages, high luminosity, large pulsation amplitude, and distinctive light curves. In this section, we explore the hypothesis that Cepheids should dominate the variability properties of a young cluster's integrated light curve. A difficulty in examining Cepheid variability with the TESS data is that the timescales of Cepheid variability tend to be much longer than those of RRL stars. As such, even short-period (4--20 day) Cepheid pulsators generally vary on timescales similar to or longer than the TESS orbit, which is similar to the timescales of the TESS systematic trends. Nevertheless, the scientific interest in identifying Cepheid variability in integrated light curves prompts us to explore what we can see of them in this pilot analysis. 

We searched the clusters in our catalog for known Cepheids. For clusters in the MCs, we cross reference known Cepheids from the OGLE catalogs \citep{soszynski_optical_2008, soszynski_optical_2010, soszynski_optical_2010-1, sokolovsky_comparative_2017}, and there were no clusters in our catalog with known Cepheids. 

\begin{figure*}[ht]
    \centering
    \includegraphics[width=.9\textwidth]{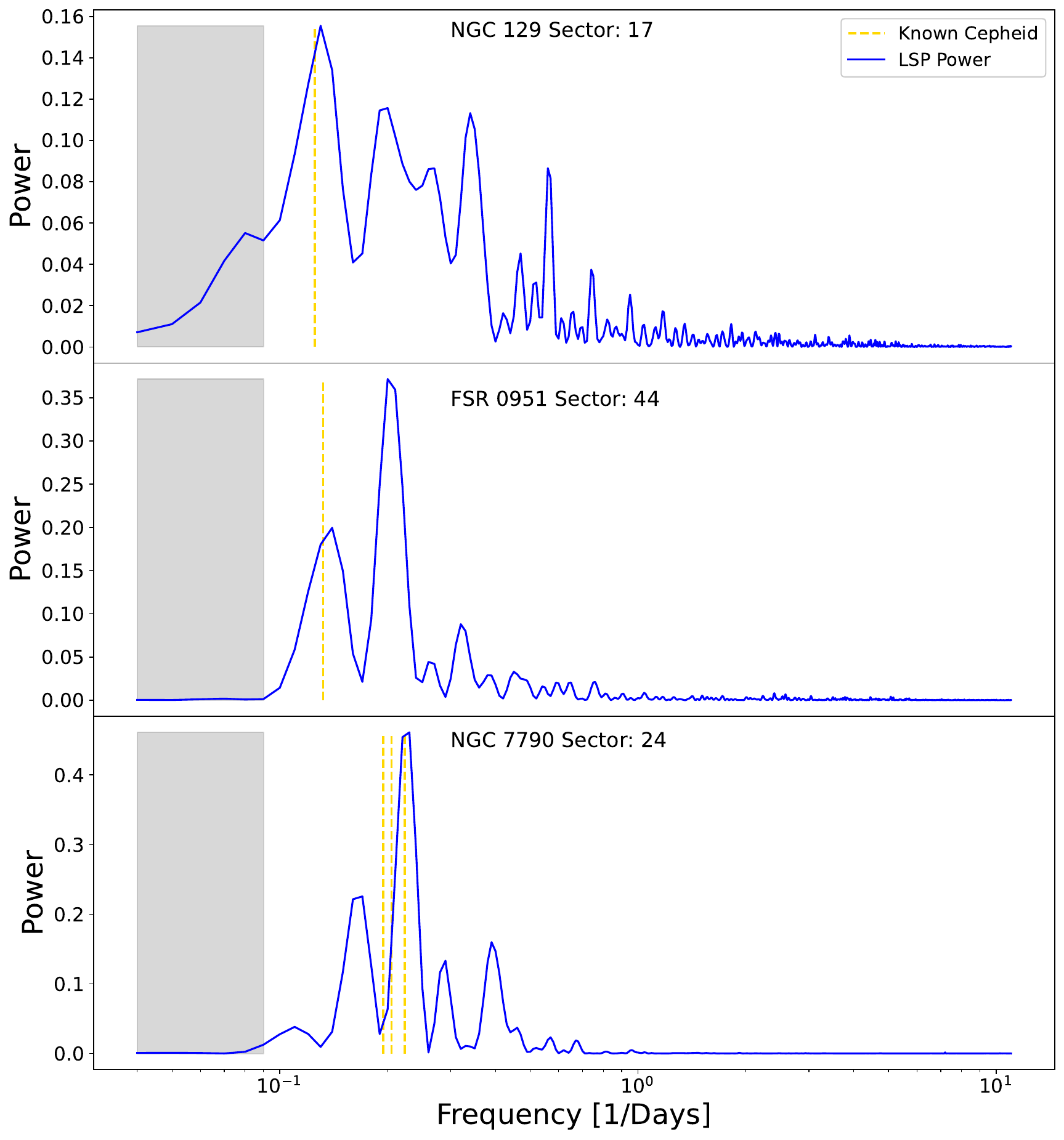}
    \caption{LSPs for thee Milky Way clusters with Cepheid variable stars inside the cluster radius \citep{medina_revisited_2021}. The blue line represents the LSP power for each cluster, and the gray region is the frequency range of the LSP to which our light curve correction methodology is not sensitive. The dashed gold lines correspond to the literature frequencies for the Cepheids. }
    \label{fig:cephs_lsp}
\end{figure*}

In the Milky Way, we select three clusters with at least one member Cepheid \citep[membership probability $>$50\%;][]{medina_revisited_2021} contained within the cluster radius, which must be smaller than 0.25$^\circ$ (Section~\ref{sec:cluster_selection}), and extract their integrated light curves (Section~\ref{sec:lce}). The LSPs for NGC~129 (one Cepheid), FSR~0951 (one Cepheid), and NGC~7790 (three Cepheids) are shown in Figure~\ref{fig:cephs_lsp}. The grayed-out region corresponds to frequencies $<$0.1~days$^{-1}$ (periods larger than 10~days), to which we are insensitive (Section~\ref{sec:lce}). The yellow dashed lines represents the literature frequency for the Cepheid(s) contained in the cluster. NGC~129 and FSR~0951 are both clusters in our catalog; however, NGC~7790 is not as it failed the compactness requirement (Section~\ref{sec:mw_clusters}). Nevertheless, we include NGC~7790 in our analysis here as it is a useful example of a cluster containing multiple Cepheids. 

For  NGC~129, there is a single sector that passed all of the quality checks from Section~\ref{sec:lcquality}, shown in the top panel of Figure~\ref{fig:cephs_lsp}. The frequency of Cepheid DL~Cas \citep[0.13~days$^{-1}$;][]{anderson_cepheids_2013} is well within the highest peak of the LSP. Using the tool described in Section~\ref{sec:lsp_maps}, we verify that the pixels associated with this peak are coming from the location of DL~Cas. 

Additionally, we test how robust this detection of a single high amplitude variable is against added noise. We perform a bootstrap re-sampling of the NGC~129 light curve (see Section~\ref{sec:lc_stats}), adding Gaussian noise to decrease the SNR, and calculate the LSP for the resulting light curve. From these noisier light curves, \texttt{elk} is able to identify a peak in the LSP at the frequency of Cepheid DL~Cas for SNRs greater than $\sim33$, compared to the SNR of $\sim6700$ reported for NGC~129.

The Cepheid RS Ori in FSR~0951 presents a more difficult case. The literature frequency of the Cepheid is 0.13~days$^{-1}$ \citep[e.g.,][]{klagyivik_observational_2009, breuval_milky_2020}. Shown in the middle panel of Figure~\ref{fig:cephs_lsp} is the cluster LSP for Sector~44, which has a peak at 0.14~days$^{-1}$, and the literature frequency of 0.13~days$^{-1}$ shown by the yellow line is clearly within this LSP peak. However, for different sectors and the sector-stitched light curve, the LSP peak is not at this period, appearing at 0.15~days$^{-1}$ (not shown in the figure). Despite the signal causing these peaks all coming from the location of RS~Ori, the light curves are not consistent across sectors, and notably in contrast with the analysis above (Section~\ref{sec:rr_ls}), increasing the time baseline by considering multiple sectors does not increase the peak strength at the literature frequency. 

However, this is not an issue unique to \texttt{elk} or our particular light curves. RS~Ori has over 18 unique extracted light curves available from TESS across these three sectors, made by different pipelines and different correction techniques \citep{kunimoto_quick-look_2021, nardiello_psf-based_2019, bouma_cluster_2019, handberg_tess_2021, caldwell_tess_2020}. Among these various light curves, we find different primary periods (as measured in the LSP). Varying levels of blending with neighboring sources -- due to different choices of effective aperture size and deblending approaches between the pipelines -- can affect the resulting LSP power coming from the Cepheid variability as well as the magnitude of TESS systematic noise that mixes with the astrophysical variability.
And in fact, RS~Ori is a heavily blended source with several bright stars within a few arc-seconds. Interestingly, the same TESS pixel containing RS Ori is also responsible for the largest peak in the LSP, at $\sim0.19$~days$^{-1}$. This periodic signal could be originating from one of the nearby stars; however, none are currently classified as variable in SIMBAD \citep{wenger_simbad_2000}. It is beyond the scope of this work to identify every source of variability within a cluster, but this case highlights one of the limitations of integrated light curves when there is (or may be) tightly packed variable sources with similar frequencies. 

NGC~7790 is widely considered the only Milky Way open cluster known to contain three Cepheids \citep{medina_revisited_2021}. Sector 24 is the only one to pass our quality checks, and its LSP is shown in the bottom panel of Figure \ref{fig:cephs_lsp}. The Cepheids CF Cas (frequency of 0.205~days$^{-1}$), CE Cas A (0.223~days$^{-1}$), and CE Cas B (0.195~days$^{-1}$) all are close in period and have similar apparent magnitudes. Furthermore, they are also all quite close on the sky, with only $\sim$2.5~arcseconds separating CE Cas A and CE Cas B \citep{sandage_double_1966} (much less than a TESS pixel), and CF Cas is located about an arcminute (roughly three TESS pixels) away. Thus, these three Cepheids are highly blended in the TESS images, and the similar periods lead to phase mixing that make it hard to individually resolve their periods in the LSP. Because there is only one sector which passes all of our quality checks, we cannot do sector stitching to mitigate this issue. Nevertheless, the LSP has its strongest peak at the frequency of CE Cas A (and thus near the frequencies of the other two). This indicates that we can partially identify individual LSP variability with specific stars even in blended TESS data. 

In summary, we extract integrated light curves for three clusters that contain Cepheid stars as probable cluster members, where the Cepheid is inside of our applied radius. We confidently recover the signals of CE Cas A and DL Cas, with a partial recovery of RS Ori and CF~Cas. Known limitations of integrated light techniques and our correction method can reasonably explain the non-detections of \textit{CE Cas B}. We note that our analysis is limited to Cepheids and RRL stars. However, clusters may have other bright, high-amplitude, periodic variables that vary on time scales shorter than a single TESS sector, such as eclipsing binaries (EBs). These types of systems should be detectable as peaks in an integrated light curve LSP and would be a useful focus of future studies.

\subsection{Low amplitude variability}
\label{sec:low_amp_variability}
The majority of clusters in our light curve catalog do not have many (or any) of the RRL or Cepheid variables discussed above. Several studies, however, have illustrated how stars in different regions of the H-R Diagram (HRD) vary with low amplitudes on different timescales \citep[e.g.,][]{mcquillan_statistics_2012, soraisam_variability_2020}. Notably, \citet{fetherolf_variability_2022} characterized the variability of over 80,000 stars observed in 2-min cadence light curves from the TESS prime mission, and for timescales less than 13 days, demonstrated drastically different typical variability periods for stars in different locations on the HRD. This large sample of stars serves as a useful testing ground for stellar variability that can be recovered in an integrated TESS light curve without high-amplitude variables. 

To characterize this variability information, we create synthetic clusters and model their integrated light curves. To perform these tests, we use the python package \textsc{SPISEA} \citep{hosek_spisea_2020} to generate MIST isochrones \citep{choi_mesa_2016} and sample them with a \citet{kroupa_variation_2001} IMF to create a synthetic star cluster of a given total mass. For each synthetic star in this cluster, we randomly select a matched star from the \citet{fetherolf_variability_2022} catalog that lies within both $5\%$ of the effective temperature and $5\%$ of the bolometric luminosity, and has a light curve best described by a sine (or double sine) function\footnote{\citet{fetherolf_variability_2022} represents the variability of each star as either a sine wave, a double sine wave, or through the auto-correlation function. They find that the stars best fit by the auto-correlation function are primarily eclipsing binaries, or high amplitude variables like Cepheids. Because this experiment is designed to characterized the variability for non-high amplitude variables, we limit the potential stellar matches to the 62,797 stars best described by a sine (or double sine) function. We note that the variability fraction varies as a function of position in the HR diagram, and stars with normalized LSP power less than 0.001 are excluded from the \citet{fetherolf_variability_2022} catalog.}.

Then, for each real star matched to an object in the synthetic cluster, we recreate the stellar light curve from the reported amplitude(s) and period(s) in \citet{fetherolf_variability_2022}. The light curve for synthetic star $i$ is modeled as one of
\begin{align}
    f_i(t) &= A_i \cos \left( \frac{2\pi t}{P_i} + \omega_i \right) \times W_{l,i} \\
    f_i(t) &= \left[A_{i,1} \cos \left( \frac{2\pi t}{P_{i,1}} + \omega_{i,1} \right)+ A_{i,2} \cos \left( \frac{2\pi t}{P_{i,2}} + \omega_{i,2} \right)\right] \times W_{l,i} \nonumber
\end{align}
depending on whether \citet{fetherolf_variability_2022} described it as a single or double sine function, where $A$ is the amplitude of the sine wave, $P$ is the period, and $f_i$ is the modeled flux of the synthetic star sampled on TESS-like time array $t$ ($0-27$~days, $\Delta t= 0.2$). To simulate the randomly offset phases of real stars, each $\omega$ is a random number between 0 and $2\pi$. $W_{l,i}$ is a weighting that represents the fractional contribution of the synthetic star to the total population luminosity. The total flux $f$ of the synthetic cluster is then the weighted sum of the individual light curves $f_i$.

\begin{figure}[t]
    \centering
    \includegraphics[width=.48\textwidth]{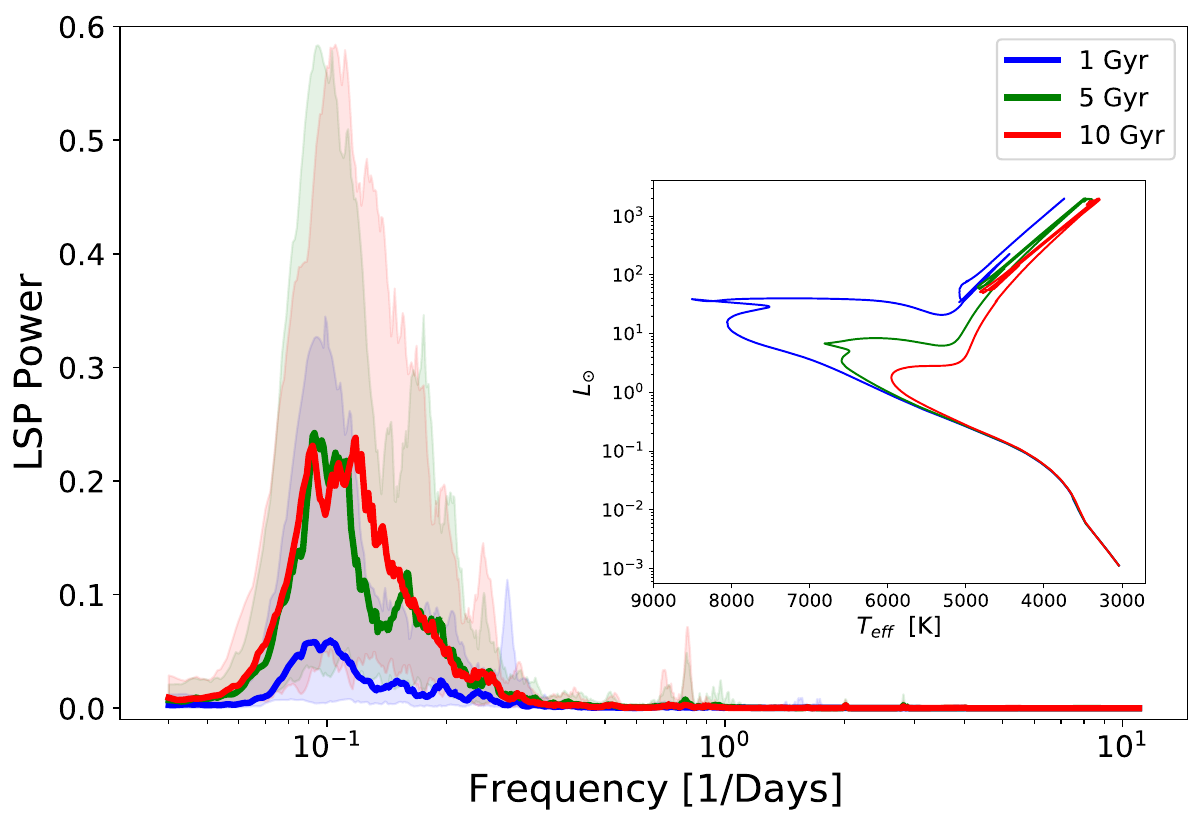}
    \caption{Shown is the LSP results for 50 trials of our generated synthetic cluster light curves for ages of 1 Gyr (blue), 5 Gyr (green), and 10 Gyr (red). Inserted are the isochrones for the 3 clusters which we then populated and matched stars from \citet{fetherolf_variability_2022}. The shaded region represents the 1$\sigma$ uncertainly of the 50 trials.}
    \label{fig:syn}
\end{figure}

We performed the steps above on isochrones with $\rm [M/H] =-0.2$ (the median metallicity of stars in the catalog) and three ages (1, 5, and 10~Gyr). Each of these isochrones is then populated for a cluster mass of $10^{3.5}$ M$_{\odot}$, as representative of physical clusters and to reduce the computing requirements. To ameliorate the effects of stochasticity and to simulate multiple observations of similar clusters, we ran 50 trials for each isochrone age and measured the LSP of the simulated integrated light curves. The results are shown in Figure~\ref{fig:syn}. The solid LSPs correspond to the median power of that age's 50 trials, the shaded regions indicate the standard deviation of the power at each frequency, and the inset plot shows the isochrones for each of the three ages.  

Given that clusters of different ages span different sections of the HR diagram, and that stars in different parts of the HR diagram tend to have different distributions of periods of variability, we may expect the LSP for the three different age samples to exhibit different LSP features. We find that the LSP's of the integrated light curves of populations with different ages do indeed look different. Most notably, the simulated 5 and 10~Gyr clusters have more power in their LSPs overall, and have different distributions of that power in frequency space. 

It has long been appreciated that the types and timescales of stellar variability change as a population ages \citep[e.g.,][]{barnes_rotational_2003, barnes_ages_2007, mcquillan_statistics_2012, mcquillan_rotation_2014, conroy_ubiquitous_2015, pinsonneault_second_2018, healy_stellar_2020, soraisam_variability_2020, fetherolf_variability_2022}. The analysis here suggests that the blended photometry of a cluster encodes information about the variability properties of individual stars. Additionally, a cluster encodes not only the variability of individual stars but also powerfully unique information about the stellar ensemble. This implies that we can use such blended time-series photometry to infer ages or other properties of the population that influence the variability of the stars. 

\subsection{Caveats}
As demonstrated in the previous sections, many different types of information are contained in integrated light curves. Here we highlight some caveats in interpreting these light curves: contamination from non-cluster member stars, phase mixing, the dominant types of stellar variability, and limitations in cadence and time span from the TESS photometry. 

First, not all the light inside our cluster aperture comes from cluster members. The presence of non-member contaminants inside of our aperture can have significant impacts on the resulting light curves. However, this issue is not unique to our analysis, and is a general, known issue with any integrated light study of clusters. Furthermore, while there have been many strides in recent years determining membership probabilities for Milky Way clusters \citep[especially using Gaia; e.g.,][]{cantat-gaudin_painting_2020,collaboration_gaia_2021, castro-ginard_hunting_2022}, these data are not available for many extragalactic samples. For any system where individual stars cannot be resolved, integrated light measurements will have to consider the impacts from non-member contaminants, likely in a statistical sense given the properties of the local field populations. 

One extreme example of field star contamination is in the cluster Collinder~236. There has long been discussion about whether the Cepheid {WZ Car} is a member of the cluster \citep[e.g.,][]{tsarevsky_search_1966, turner_does_2009}, though more recently \citet[][]{turner_does_2009} determined that the cluster is nearly two times closer to Earth than WZ Car, even though visually they are located in the same place on the sky. However, while the Cepheid is not a member of the cluster, it is the brightest star in the aperture, dominates the integrated flux, and produces the strongest peak in the LSP. This example illustrates the need for future work on addressing the effects of flux contamination in integrated light curves. 

Second, when considering integrated light curves with blended sources, phase mixing will be a prominent issue. However, as demonstrated in Section~\ref{sec:rr_ls}, this issue can be mitigated by considering longer time baselines, such as stitching multiple TESS sectors together (but see also Section~\ref{sec:cepheids}). Such approaches, however, will not disentangle the effects of phase mixing if the periodic variations migrate in phase (such as spot modulations due to rotation).

Third, all integrated light curves will be most impacted by stars that are significantly brighter or those that produce greater variability than other sources in the population. Low-amplitude variability of especially bright stars can appear prominently in a LSP, while a relatively fainter variable can also impact the LSP if its amplitude is especially large. In general, signals in integrated light curves will be generated by a product of stellar brightness and amplitude of variation.

Lastly, the telescope and survey properties will deeply impact the utility of integrated light photometry. The optical setup, including aperture, pixel scale, exposure time, and field of view, will limit the populations where this overall approach can be applied. The observing characteristics, such as cadence, duty cycle, time baseline, and total number of observations will affect the types of variability that can be recovered by limiting the amplitudes and periods of variation that can be captured.

\section{Summary}
\label{sec:sum}

We have computed the first TESS-based integrated light curves of star clusters. Our methodology is implemented in the open source package \texttt{elk}, which uses principal component analysis and TESS systematic corrections to extract accurate light curves within a given aperture. Our light curves for 348 clusters in the Milky Way, Small Magellanic Cloud, and Large Magellanic Cloud are available through MAST as a High Level Science Product via \dataset[10.17909/
]{\doi{ 10.17909/f4kb-vc30 }}.
We confirm that our star cluster light curves preserve the stellar variability information for a number of stars in the cluster, both high-amplitude variables and low-amplitude variations. 

Because the information from individual stars is preserved in the ensemble, the light curves presented here can be used to probe stellar astrophysics. This includes building a comprehensive sample of the variability characteristics of stellar populations, spanning a large range of parameters like metallicities, masses and ages, to better understand the types of stellar variability and the connection between integrated light variability and population parameters.

\begin{acknowledgments}
TMW, GZ, and VJP acknowledge support by the Heising-Simons Foundation through grant 2019-1492. JP, MPC, JTD, and AP acknowledge support by the Heising-Simons Foundation through grant 2019-1491.
This work was supported the National Science Foundation under Grant No. 1852010.
T.F. acknowledges support from the University of California President's Postdoctoral Fellowship Program.

This paper includes data collected with the TESS mission, obtained from the MAST data archive at the Space Telescope Science Institute (STScI). Funding for the TESS mission is provided by the NASA's Science Mission Directorate. STScI is operated by the Association of Universities for Research in Astronomy, Inc., under NASA contract NAS 5–26555.
This work also uses data from the European Space Agency (ESA) space mission Gaia. Gaia data are being processed by the Gaia Data Processing and Analysis Consortium (DPAC). Funding for the DPAC is provided by national institutions, in particular the institutions participating in the Gaia MultiLateral Agreement (MLA). The Gaia mission website is https://www.cosmos.esa.int/gaia. The Gaia archive website is https://archives.esac.esa.int/gaia.

This research made use of \textsc{lightkurve}, a Python package for Kepler and TESS data analysis (Lightkurve Collaboration, 2018).

We would also like to recognize and thank Samantha-Lynn Martinez\footnote{Her porfolio can be found at \url{https://www.samanthalynnmartinez.com/illustrationdesign}} for designing the \texttt{elk} logo.
\end{acknowledgments}

\facilities{TESS, Gaia}

\software{Astropy \citep{astropy:2013, astropy:2018, astropy:2022}, Lightkurve \citep{lightkurve_collaboration_lightkurve_2018}, astroquery \citep{astroquery_2019}}
Matplotlib \citep{Hunter2007}, NumPy \citep{Harris2020}, SciPy \citep{Virtanen2020}
%

\bibliographystyle{aasjournal} 

\bibliography{references} 

\end{document}